\documentclass[aps,prl,twocolumn,superscriptaddress]{revtex4-2}

\usepackage{graphicx}
\usepackage[squaren]{SIunits}
\usepackage{amsmath,amsthm,amssymb}
\usepackage{wasysym}
\usepackage[colorlinks=true,allcolors=blue]{hyperref}
\usepackage[normalem]{ulem}

\renewcommand{\d}{\mathrm{d}}
\newcommand{\D}{\mathrm{D}}
\usepackage{bm}

\begin{document}

\title{Adhesion differentials control the rheology of biomimetic emulsions}
\author{Quentin Guigue}
\author{Marc Besse}
\author{Raphael Voituriez}
\author{Alexis M. Prevost}
\author{Elie Wandersman}
\affiliation{Sorbonne Universit\'e, CNRS, Laboratoire Jean Perrin, , LJP, F-75005 Paris, France}
\affiliation{Sorbonne Universit\'e, CNRS, Inserm, Institut de Biologie Paris-Seine, IBPS, F75005 Paris, France}

\author{Matthias Merkel}
\email[Contact author: ]{matthias.merkel\@univ-amu.fr}
\affiliation{Aix Marseille Univ, Universit\'e de Toulon, CNRS, CPT (UMR 7332), Turing Center for Living Systems, 13009 Marseille, France}

\author{Lea-Laetitia Pontani}
\email[Contact author: ]{lea-laetitia.pontani\@sorbonne-universite.fr}
\affiliation{Sorbonne Universit\'e, CNRS, Laboratoire Jean Perrin, , LJP, F-75005 Paris, France}
\affiliation{Sorbonne Universit\'e, CNRS, Inserm, Institut de Biologie Paris-Seine, IBPS, F75005 Paris, France}

\date{\today}

\begin{abstract}
Animal morphogenesis involves complex tissue deformation processes, which require tight control over tissue rheology.
Yet, it remains insufficiently understood how tissue rheology results from the interplay between cellular packing and cellular forces, such as cortical tension, cell pressure, and cell-cell adhesion.
Here, we follow a biomimetic approach to study this interplay.
We mimic adhesive cells with oil droplets whose adhesion strength and specificity can be flexibly tuned.
Using microfluidics, we expose 2D emulsions to an oscillatory geometry imposing cyclic pure shear, and we develop a geometric method to quantify their rheology using only imaging data.
We find that some of the emulsions made of two droplet types progressively change their yielding behavior across subsequent shear cycles.
Combining this with vertex model simulations, we show that the observed shift in yielding behavior is due to a progressive compaction, which only occurs in emulsions with a high adhesion differential and only when exposed to oscillatory shear.
Gradients of cell compaction have been observed during animal development. Our work demonstrates how such gradients can be used to control gradients of tissue rheological properties.
Moreover, the progressive compaction suggests the emergence of a pumping mechanism, which potentially acts in many cellular materials, from foams to tissues.
\end{abstract}


\maketitle

\section{Introduction}

In broad strokes, animal development consists of patterning, i.e.\ ensuring cells take on the correct biochemical identity in the right places, and morphogenesis, i.e.\ ensuring tissues deform to obtain their correct adult shapes \cite{Wolpert2015}.
Changes of tissue shape are driven by internal and external forces \cite{Goodwin2021}, but the way these forces translate into tissue deformation crucially depends on the tissue mechanical properties, i.e.\ tissue rheology.
Tissue rheology depends in turn on cellular forces \cite{Noll2017, Mongera2018, Petridou2019, Montel2022, Dessalles2024}, cell mechanical properties \cite{Bi2015, Stirbat2013, Sadeghipour2018, DAngelo2019} and on the tissue structure \cite{Mongera2018,Petridou2021,Michaut2025}, i.e.\ the way cells are packed within the tissue.
Yet, these properties are modulated by the biochemical identities of the involved cells \cite{Kashkooli2021}.
Moreover, whenever forces and deformations are applied to tissues, they could in turn modify the cellular packing.
How all these effects combine to affect tissue rheology and thus morphogenesis is still not fully understood.

A cell-scale force that is particularly important for tissue function and development overall is cell-cell adhesion.
Specifically, it has been shown to play a key role in the patterning of germ layers \cite{Townes1955,Schoetz2008}, in shape changes during gastrulation \cite{Morita2017,Wallmeyer2018}, in the emergence of epithelial cell polarity \cite{Rustarazo-Calvo2025}, and it could play a role during vertebrate axis formation \cite{Gsell2023}.
Indeed, synthetic biological systems explicitly demonstrated the capacity of heterogeneous adhesion to separate cell populations \cite{Cachat2016,Courte2024}.
From a theoretical perspective, understanding such cell population separation by heterotypic adhesion has first been discussed by Steinberg \cite{Steinberg1963,Foty2005}, where adhesion was proposed to reduce the effective interface tension between tissues.
Yet, later studies showed that cell-cell adhesion molecules also affect cortical contractility, substantially amplifying the effect that adhesion has on effective interface tensions \cite{Brodland2002,Manning2010a,Maitre2012,Amack2012}.
Moreover, models for biological tissues, such as the vertex model, the Potts model, or continuum models, often describe the effect of adhesion by reducing cell-cell interface tensions \cite{Graner1992,Manning2010a,Gsell2022}. The effect of adhesion can also be captured differently, for instance by means of a higher energy barrier towards separating individual cells \cite{Maitre2012,Pawlizak2015}, which would also lead to a higher barrier towards cell rearrangements within a tissue.
Such barriers provide another way in which cell adhesion can affect tissue rheology \cite{Erdemci-Tandogan2021a}.
Taken together, while the role of differential adhesion for the separation of cell populations and the related tissue surface tension has been firmly established, a direct effect of differential adhesion on tissue bulk rheology has never been reported to the best of our knowledge.

\begin{figure*}[t]
\centering
\includegraphics[width=\textwidth]{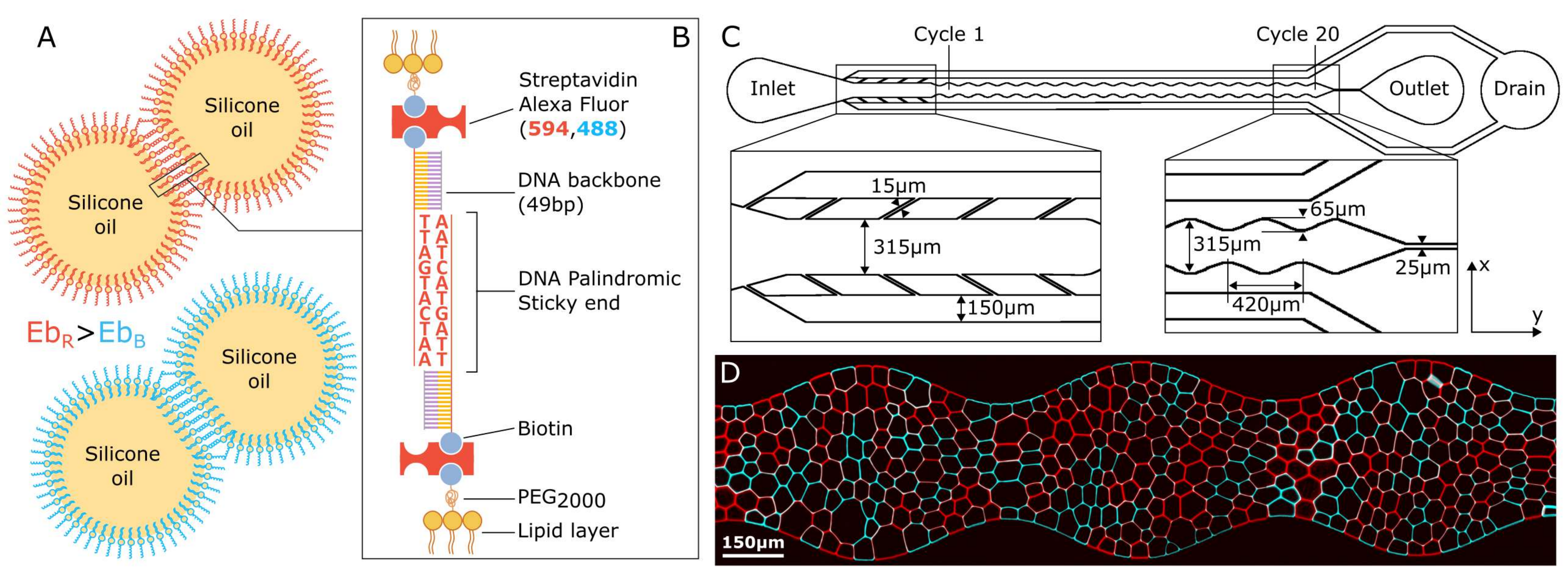}
\caption{(A) Schematic representation of DNA-functionalized droplets. Silicone oil droplets are stabilized with SDS, EPC and biotinylated phospholipids. Two droplets holding the same DNA strand will interact together upon contact with a given binding energy $Eb$, while there is no adhesion between droplets holding different strands. The average diameter of these functionalized droplets is $29.6\,\micro\meter$ (B) Biotinylated palindromic DNA strands are grafted on the lipids hydrophilic heads using a streptavidin bridge. Alexa Fluor 594 (red) or 488 (blue) is attached to the streptavidin, allowing to identify which DNA strand a given droplet carries. (C) 2D representation of the microfluidic channel that was designed to apply periodic shear deformation to the emulsions. The channel height is $30\,\micro\meter$, which is adapted to the droplets diameter. Note that the straight channel used for control experiments has the same design, except that the wavy borders are replaced by purely straight wall. (D) Stitched confocal image of a P0/P10 emulsion in the wavy microfluidic channel (from left to right, undulations \#11 to \#13; blue is P0, red is P10).}
\label{figure1}
\end{figure*}

Here we study how cell-cell adhesion affects tissue rheology using a well-controlled biomimetic system.  
We use aqueous emulsions of oil droplets whose adhesions can be tuned in order to mimic the minimal adhesive and mechanical properties of cells in soft tissues, as first described in \cite{Pontani2012}. This model system allows to isolate the role of adhesion from many other parameters present in biological tissues. 
In previous studies, we used such biomimetic systems to show how adhesion tunes the material response to a single mechanical perturbation \cite{Golovkova2020, Golovkova2021}. In particular, we showed that droplet rearrangements were slower and spatially delayed in the presence of adhesion, causing an increase of oriented elastic droplet deformations.
In static assemblies of droplets, we also showed how the percolation of the adhesive contacts network could control the packing topology and large-scale deformations~\cite{Montel2021}.
However, so far, the role of heterogeneous adhesion on the tissue-scale rheology in such biomimetic systems remains unclear.

Here, we introduce different droplet types with differential adhesion and study how it affects the yielding behavior of the emulsion.
We introduce differential adhesion through distinct DNA binders on two droplet types.
We probe the yielding behavior of these emulsions by flowing them through an undulated 2D microfluidic channel in order to apply oscillatory pure shear deformations.
Surprisingly, we find that high adhesion differentials lead to an increase of droplet shape anisotropy across subsequent shear cycles. 
To understand these observations, we decompose the applied shear into contributions by droplet shape change and droplet rearrangements \cite{Merkel2017}.
Here, we extend this method, defining a \textit{reversible fraction} $f_r$, which is the fraction of the overall shear that is created by droplet shape changes.
We show that $f_r$ depends mainly on the local droplet shape anisotropy. This relationship can act as a rheological constitutive relation, capturing most of the yielding behavior of the emulsion, which we demonstrate by predicting the observed droplet shape variations.
Guided by simulations of a cell-based model, we could show that (i) the yielding behavior of the emulsions strongly depends on the droplet packing fraction, and (ii) the emulsions with heterogeneous adhesions show a progressive compaction in terms of an increase of droplet packing fraction across subsequent shear cycles.
Yet, this compaction occurred exclusively for emulsions with heterogeneous adhesion and only when an oscillatory shear is applied.
Taken together, our findings show that an adhesion differential modifies the flow properties of adhesive emulsions, leading to their compaction under repeated shear deformations, an associated shift in their rheological properties, and an increase of droplet shape anisotropy.
We expect our work to help better understand dynamic changes that biological tissues undergo during morphogenesis. 
This includes for instance the formation of packing fraction gradients \cite{Mongera2018,Petridou2021,Michaut2025}, and mechanisms creating hydraulic extra-cellular flows \cite{Dumortier2019}.

\section{Results}

\subsection*{Introduction of an adhesion hierarchy into biomimetic emulsions}

We use biomimetic emulsions that are made of athermal oil droplets dispersed in an aqueous solution \cite{Pontani2012,Feng2013, Pontani2016}. 
To control droplet-droplet adhesion, DNA strands are grafted onto the droplet surfaces through streptavidin bridges~\cite{Pinon2018,Zhang2017} (see Fig.~\ref{figure1}A,B and Materials and Methods).
Different droplet populations are created by using different DNA sequences, which are distinguished by fluorescently labeling the streptavidin bridges with different colors.
Our DNA constructs are all made of a double-stranded backbone of 48 base pairs (bp) followed by a single passive base serving as a flexible junction, and a single strand, the ``sticky end'' (see Fig.~\ref{figure1}A,B, SI for the full sequences). 
In this work we use sticky ends with palindromic sequences, i.e.\ sequences that bind to themselves. 
The binding energy of the sticky ends increases with their length~\cite{SantaLucia1998}, and their sequences are chosen such that two DNA strands with different sequences should not bind to each other. 
This allows us to mix together two populations of droplets that are self-adhesive but do not exhibit any cross adhesion.

Specifically, we work with four distinct constructs: the P0 construct only contains the double stranded backbone in the DNA sequence, making these droplets non-adhesive; the P6, P10 and P14 constructs are made with sticky ends of lengths 6, 10, and 14~bp, respectively. We hereafter use these construct names to refer to the correspondingly functionalized emulsions. We studied homogeneous emulsions (adhesive P10 emulsions or non-adhesive P0 ones), but also mixed emulsions such as P0/P10, P6/P10, P0/P14 and P6/P14, with a 1:1 volume ratio, in order to probe various adhesion hierarchies. 
With these mixtures we vary two parameters independently of each other. The average binding energy in the emulsion depends on the sum of the binding energies associated with each droplet type. In that sense, the P0/P10 mix exhibits a lower average binding energy than the P0/P14 or P6/P10 mixes. Alternatively, one can take into account the difference between the binding strength of the two droplet types, e.g.\ the adhesion differential is larger for P0/P14 and P0/P10 mixtures as compared to the P6/P10 one.

We flow these emulsions in microfluidic channels that are designed to apply repetitive pure shear deformations to the system (see Fig.~\ref{figure1}C,D). 
The total amplitude of each oscillation, which corresponds to a total strain amplitude of $\sim50\%$, is sufficient to induce plastic rearrangements (see Video S1 and Fig~\ref{fig:Fig3}B) \cite{Golovkova2020}. 
We flow the emulsions sufficiently slowly to ensure that the droplets can re-adhere between constrictions (see Materials and Methods). 
Finally, the emulsions are imaged in two ways: either the flow is stopped and static images are acquired at all undulations, or the flow is maintained and a movie is acquired at a given undulation in order to track droplet dynamics and rearrangements (see Materials and Methods). 

We first seek to understand if there is an effect of adhesion on the structure of the emulsion throughout the channel. To do so, we measure the number and total length of homotypic (red/red or blue/blue) and heterotypic (red/blue) contacts between neighboring droplets on static images. We find that, even with the largest adhesion differentials, the length proportion of heterotypic contacts does not change significantly as the emulsion progresses through the channel (see Figure~S4 right). This indicates that segregation between the two droplet populations does not take place over the twenty shear cycles of the channel. However, significant effects are unveiled regarding droplet shapes, as discussed in the next section.

\begin{figure*}[t]
\centering
\includegraphics[width=0.95
\textwidth]{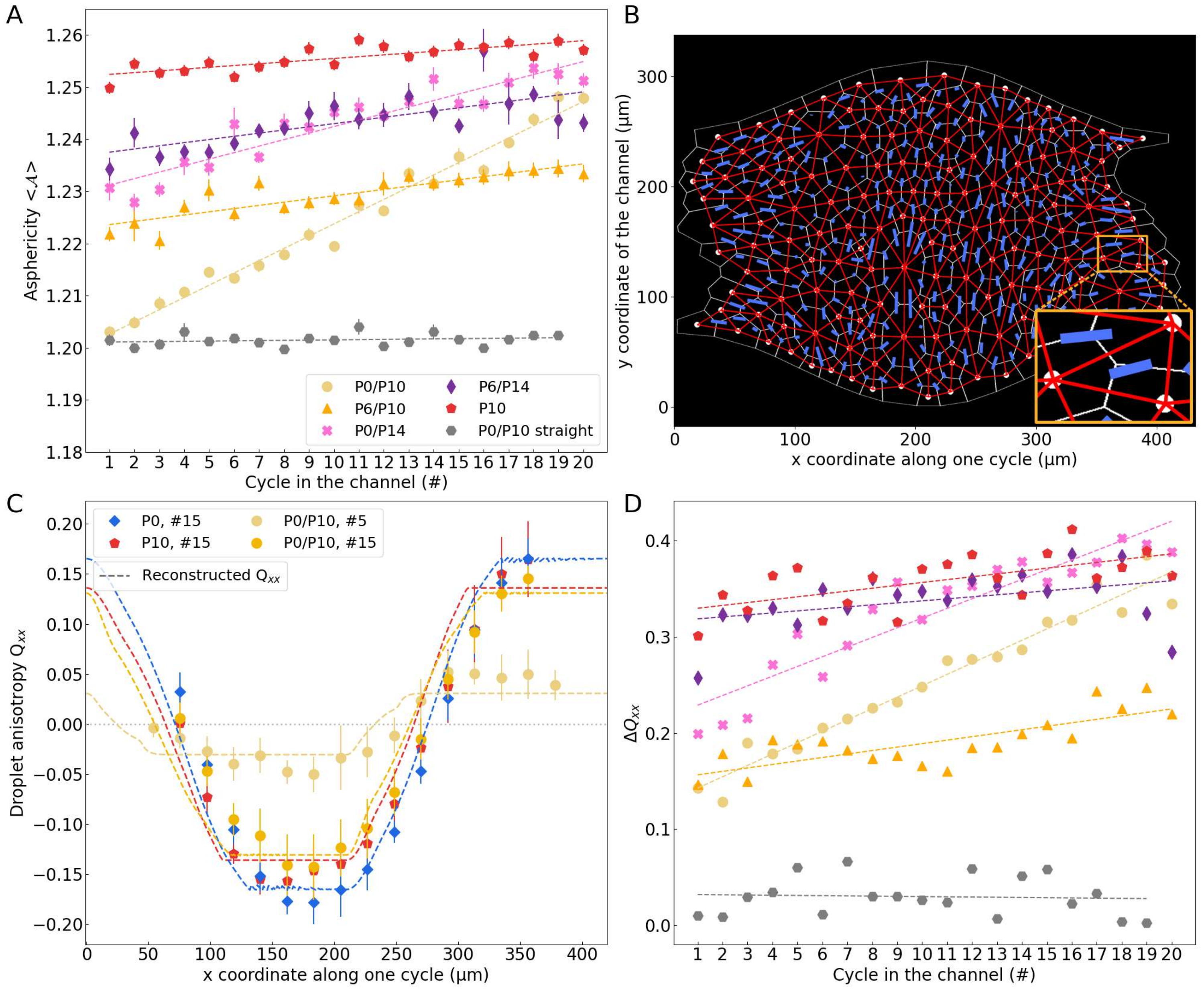}
\caption{(A) Average asphericity $\langle \mathcal{A} \rangle$ computed over all the droplets in each undulation cycle in the channel for different conditions (P0/P10 emulsions: yellow circles, P6/P10: orange triangles; P0/P14: pink crosses; P6/P14: purple thin diamonds; P10: red pentagons) or at equivalent cycles in the straight channel (grey circles). The analysis was performed on static images acquired after the flow was arrested in the channel. Error bars represent the standard error of the mean across experimental repetitions. Each data point was averaged over about 1500 droplets (min=993, max=2233, average=1488). (B) Example of a triangulated droplet network (red) overlaid on the image of a segmented emulsion. Blue bars represent the orientation and magnitude of $\bm{Q}$, computed for each triangle (see magnified inset). (C) Evolution of $Q_{xx}$ along the x-axis for undulations \#5 (yellow circles) and \#15 (gold circles) of a P0/P10 heterogeneous emulsion as well as undulation \#15 for P10 (red pentagons) and P0 (blue diamonds) homogeneous emulsions in the oscillatory channel. Values are measured in movies and averaged over time, error bars represent the standard deviation. Dashed lines represent the prediction of $Q_{xx}(x)$, reconstructed for each condition from the reversible fraction and the observed velocity field in the movie. (D) Evolution of $\Delta Q_{xx}$, calculated as the amplitude of $Q_{xx}(x)$ variations within a channel oscillation, as a function of the undulation for each experimental condition (same color code as in panel A). Similarly to the results displayed in panel (A), P0/P10 (yellow) and P0/P14 (pink) emulsion exhibit a specific linear increase, whereas other conditions display more stable $\Delta Q_{xx}$ values along the channel. 
In the case of the P0/P10 in the straight channel (gray), $\Delta Q_{xx}$ is constant near 0 as there is no geometrical constraint inducing any ordered anisotropy in the emulsion.}
\label{fig:shape}
\end{figure*}

\subsection*{Increase of droplet shape anisotropy along the channel}
We explore droplet shape for different adhesion configurations on static acquisitions in the successive undulations. Indeed, adhesive emulsions flowing in constrictions are expected to be deformed as their plastic response is impaired by droplet-droplet adhesion \cite{Golovkova2020, Golovkova2021}. 
Here we study this effect in the presence of repeated shear and heterogeneous adhesion.
To do so, we measure the asphericity $\mathcal{A} = p^2/4 \pi a$, which compares the perimeter $p$ of a droplet to its cross-sectional area $a$. It is one for a circular disc, and increases as the droplet shape deviates from that of a circle. We plot the asphericity averaged over all droplets in each undulation cycle of the channel as a function of the cycle number in the channel and for all adhesion conditions (Fig.~\ref{fig:shape}A). 
Some conditions exhibit a larger linear increase of droplet asphericity as the emulsion progresses through the channel. Notably, the sharpest increases are observed for emulsions that exhibit high adhesion differentials between droplet populations such as P0/P10 (yellow circles) and P0/P14 (pink crosses). 
This indicates that this increase in droplet asphericity under repeated shear depends on the adhesion differential in the system. Strikingly, the deformation value associated with the lowest average binding energy, i.e. the P0/P10 condition, even surpasses the one of the P6/P10 condition after the thirteenth cycle in the channel (Fig.~\ref{fig:shape}A). This suggests that the presence of an adhesion energy differential can be more important than the average adhesion energy in creating droplet asphericity.

This is in contrast with previous descriptions of static adhesive droplet packings in which the equilibrium shape of a droplet is given by the balance between the binding energy gain and the energetic cost of surface deformation due to the oil/water surface tension \cite{Pontani2012}. According to these simple considerations, a stronger binding energy automatically induces larger droplet deformations. Here the rheology of such adhesive emulsions under repeated shear seems incompatible with this static vision.
In order to verify if this increase is indeed due to the applied shear, we use a control experiment in which a P0/P10 mixture experiences a plug flow inside a straight channel with otherwise similar dimensions (see SI). In this case, the average droplet shape remains constant throughout the whole channel (Fig.~\ref{fig:shape}A, grey points), confirming the central role of repeated shear deformations for the increase of droplet asphericity. 

\begin{figure}[th!]
%\centering
\includegraphics[width=0.364\textwidth]{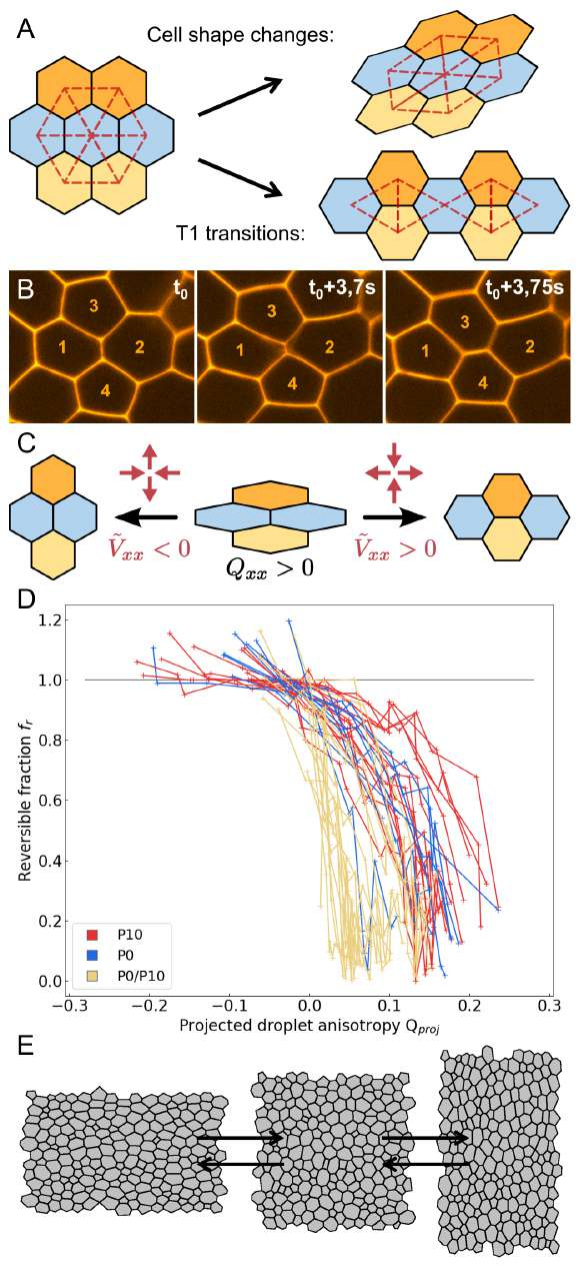}
\caption{(A) The observed shear is decomposed into a contribution from cell shape changes (elastic deformation, reversible, upper right sketch) and a contribution from T1 transitions (plastic events, irreversible, lower right sketch). 
(B) Confocal images of a T1 transition in a P10 emulsion. Droplets 1 and 2, initially neighbors at $t_{0}$, are progressively pulled apart ($t_{0}+3.7s$), until they are no longer in contact, letting droplets 3 and 4 become new neighbors ($t_{0}+3.75s$).
(C) Whether and how much shear is accommodated elastically vs.\ plastically depends on its direction. If shear is oriented perpendicular to droplet reorientation (left arrow) one expects purely elastic deformation. If shear is oriented parallel to droplet shape, a part of the shear may be accommodated by plastic rearrangements.
(D) Reversible fraction as a function of the projected elongation $Q_\mathrm{proj}=\mathrm{sgn}(\tilde{V}_{xx})\, Q_{xx}$. All curves are labeled according to the experimental conditions (red: P10 emulsions, blue: P0 emulsions, yellow: P0/P10 emulsions), independently of channel undulation. Each curve represents the analysis of one movie.
(E) Schematic of vertex model simulations. We use periodic boundary conditions, whose dimensions are modified to impose oscillatory pure shear deformations at a constant total area.}
\label{fig:Fig3}
\end{figure}

In order to understand how channel geometry may affect droplet shape, we quantify not only the magnitude, but also the orientation of droplet shape anisotropy. Indeed, previous work on foams \cite{Graner2008, Dollet2015}, emulsions \cite{Golovkova2020, Golovkova2021} and tissues \cite{Blanchard2009, Etournay2015, Guirao2015, Merkel2017, Tlili2022, Butler2009} related the droplet or cell shape and its orientation to the overall material shear deformation.
Here, following Refs.~\cite{Etournay2015, Merkel2017}, we probe local droplet elongation using a symmetric, traceless tensor $\bm{Q}$, which quantifies both magnitude and orientation of droplet shape anisotropy. To this end, we first triangulate the emulsion by connecting the centers of neighboring droplets by triangles (Fig.~\ref{fig:shape}B inset), leading to a triangulation of the whole emulsion without gaps or overlaps (Fig.~\ref{fig:shape}B, details in SI). For a given triangle, the magnitude of $\bm{Q}$ corresponds to $\log(\mathrm{AR})/2$, where AR is the aspect ratio of an ellipse fitted to the triangle, and the orientation of the tensor $\bm{Q}$ corresponds to that of the long axis of the ellipse. For any given region of the emulsion, we define the local droplet shape as the area-weighted average of the triangle-based $\bm{Q}$ tensors (details in the SI).

In Fig.~\ref{fig:shape}C, we show examples of time-averaged $Q_{xx}$, which is the horizontal component of $\bm{Q}$, versus the position $x$ along the channel within one undulation, averaged across the channel width $y$ (see Fig.~\ref{fig:shape}B). A positive horizontal droplet shape component, $Q_{xx}>0$, indicates a droplet elongated along the channel direction, whereas a negative component, $Q_{xx}<0$, indicates a droplet elongated perpendicular to the channel direction. We find that $Q_{xx}$ oscillates with positive values, i.e.\ horizontally aligned droplets, where the channel is narrow, and negative values, i.e.\ vertically aligned droplets, where the channel is wide (Fig.~\ref{fig:shape}B,C).
We furthermore quantified $\Delta Q_{xx}:=Q_{xx}^\mathrm{max}-Q_{xx}^\mathrm{min}$, the difference between maximal and minimal values of $Q_{xx}$ within a single channel undulation (Fig.~\ref{fig:shape}D), and find that it behaves very similarly to the asphericity $\mathcal{A}$ (Fig.~\ref{fig:shape}A).
Specifically, we observe a substantial shift for emulsions with a large adhesion differential, such as P0/P10 (Fig.~\ref{fig:shape}A,D).

\subsection*{Geometric analysis of the yielding behavior}
To understand the evolution of the droplet shape anisotropy $Q_{xx}$ for different emulsions both with and without oscillatory shear, we study the yielding behavior of these emulsions. 
To this end, we use the fact that any anisotropic deformation of a cellular material can be decomposed into contributions by cell shape changes and by cell rearrangements, so-called T1 transitions (Fig.~\ref{fig:Fig3}A,B).
Specifically, for a given set of droplets that we experimentally track over time, we use the formalism from Ref.~\cite{Merkel2017} to decompose the rate tensor (i.e.\ the anisotropic part of the strain rate tensor) $\bm{\tilde{V}}$ as follows:
\begin{equation}
  \bm{\tilde{V}} = \frac{\D\bm{Q}}{\D t} + \bm{R}. \label{eq:shear-rate-decomposition}
\end{equation}
with $\bm{\tilde{V}}$ being the shear rate tensor averaged over the tracked region, which is decomposed into the rate of change of the average droplet shape anisotropy $\bm{Q}$, where $\D/\D t$ is an \textit{advective, corotational} derivative, and a contribution from T1 transitions $\bm{R}$ occurring within this region (Fig.~\ref{fig:Fig3}A).
In other words, the first term on the right-hand side in \eqref{eq:shear-rate-decomposition} represents reversible, elastic contributions to shear, while the second term represents irreversible, plastic contributions.
Note that we slightly simplify \eqref{eq:shear-rate-decomposition} as compared to Ref.~\cite{Merkel2017}. First, we neglected contributions that appear only in biological tissues, such as cell divisions. Second, for simplicity, we choose a quasi-1D description, neglecting the corotational contribution, and focusing only on the components along the channel direction:
\begin{equation}
  \tilde{V}_{xx} = \frac{\d Q_{xx}}{\d t} + R_{xx}, \label{eq:shear-rate-decomposition_xx}
\end{equation}
where $\d/\d t=\partial Q_{xx}/\partial t + \mathrm{v}_x(\partial Q_{xx}/\partial x)$ denotes the advective derivative, with $\mathrm{v}_x$ being the $x$ component of the local velocity.
Note that in our emulsions, we can measure all terms in Eqs.~(\ref{eq:shear-rate-decomposition}) and (\ref{eq:shear-rate-decomposition_xx}) (see SI, Figure~S7).

The mechanical relaxation rate of our emulsions is much faster than the shear rate that we impose through the undulations, i.e.\ the emulsions are deformed quasi-statically (see SI).
This condition of quasi-static deformation allows us to go beyond the formalism of Ref.~\cite{Merkel2017}. Since the shear time scale does not play any role, we divide \eqref{eq:shear-rate-decomposition_xx} by  $\tilde{V}_{xx}$:
\begin{equation}
  1 = f_r + f_i, \label{eq:reversible fraction}
\end{equation} 
where $f_r = (\d Q_{xx}/\d t)/\tilde{V}_{xx}$ is the fraction of the overall shear that is due to reversible (i.e.\ elastic) droplet shape deformations, and $f_i = R_{xx}/\tilde{V}_{xx}$ is the fraction of the overall shear that is due to irreversible (i.e.\ plastic) droplet rearrangements.
If we can understand for each of our emulsions what fraction of the applied strain changes droplet shapes, $f_r$, vs.\ what fraction is accommodated by droplet rearrangements, $f_i=1-f_r$, we will be able to predict droplet anisotropy $Q_{xx}$ as the emulsion is pushed through the undulating channel. 
In other words, knowing the behavior of $f_r$ amounts to knowing a \emph{constitutive relation}. 
Note that, while the stress tensor does not explicitly appear in these equations, it can be easily computed from the droplet shape anisotropy and the interface tensions using the Bachelor formula \cite{Batchelor1970,Kraynik2003,kabla2007quasistatic}.

In general, the reversible fraction $f_r$ will depend on the state of the emulsion, particularly on the average droplet shape anisotropy $Q_{xx}$.
Yet, it likely also depends on whether the emulsion is sheared parallel to the droplet elongation axis, i.e.\ $\tilde{V}_{xx}Q_{xx}>0$, or perpendicular to the droplet elongation axis, i.e.\ $\tilde{V}_{xx}Q_{xx}<0$ (Fig.~\ref{fig:Fig3}C). 
Specifically, in a simple picture, one would expect that if the emulsion is being sheared \textit{parallel} to $\bm{Q}$ (Fig.~\ref{fig:Fig3}C, right arrow), this would in part increase droplet elongation and in part lead to droplet rearrangements. Meanwhile, when the emulsion is sheared \textit{perpendicular} to $\bm{Q}$ (Fig.~\ref{fig:Fig3}C, left arrow), one would just expect that the droplet shape relaxes towards a less elongated state, while almost no droplet rearrangements are expected to occur.
Thus, we expect $f_r$ to depend mostly on the projection of $Q_{xx}$ on the shear direction, $Q_\mathrm{proj}:=\mathrm{sgn}(\tilde{V}_{xx})\, Q_{xx}$, where $\mathrm{sgn}(\tilde{V}_{xx})$ denotes the sign of $\tilde{V}_{xx}$.

To measure how the reversible fraction $f_r$ depends on the projected droplet shape $Q_\mathrm{proj}$ for different emulsions, we acquired movies for conditions that exhibit the most salient differences: non-adhesive P0 homogeneous emulsions, adhesive P10 homogeneous emulsions, and P0/P10 mixtures of droplets.
The movies were acquired in different undulations of the channel (\#5, \#10 and \#15). Note that droplets in these experiments are all labelled with the same fluorophore, even in the case of heterogeneous P0/P10 emulsions, in order to acquire the movies at a high enough frame rate so that individual droplets can be tracked.
In order to measure the reversible fraction function $f_r(Q_\mathrm{proj})$ in these movies, we essentially divide each image into stripes across the whole channel height, and with a width of $96\,\mathrm{pixels}$ ($\simeq 21.5\,\micro\meter$).
For each stripe, we quantify the average triangle elongation $Q_{xx}$ and the shear rate $\tilde{V}_{xx}$, which we average over all time points (see SI for details). We also check that for any given position, droplet shapes are essentially stationary over time (see SI, green curve in Figure~S7). We then compute the projected shape and reversible fraction for that stripe as $Q_\mathrm{proj}=\mathrm{sgn}(\langle\tilde{V}_{xx}\rangle_t)\,\langle Q_{xx}\rangle_t$ and $f_r=\langle\d Q_{xx}/\d t\rangle_t/\langle\tilde{V}_{xx}\rangle_t$, respectively, where $\langle\cdot\rangle_t$ denotes an average over time. 

The resulting $f_r(Q_\mathrm{proj})$ curves for all experimental conditions are presented in Fig.~\ref{fig:Fig3}D, in which we pooled under the same label all the movies acquired at all undulations for one type of emulsion. The curves associated to the P0/P10 emulsions appear shifted to the left with respect to the homogeneous P0 and P10 emulsions: at a fixed value of $f_r < 1$, these emulsions yield at a lower value of $Q_\mathrm{proj}$. In other words, in order for them to yield, the P0/P10 emulsions do not need to be deformed much, i.e.\ one does not need to elongate the droplets very much until they start to rearrange. 

\begin{figure*}[ph!]
%\centering
\includegraphics[width=0.9\textwidth]{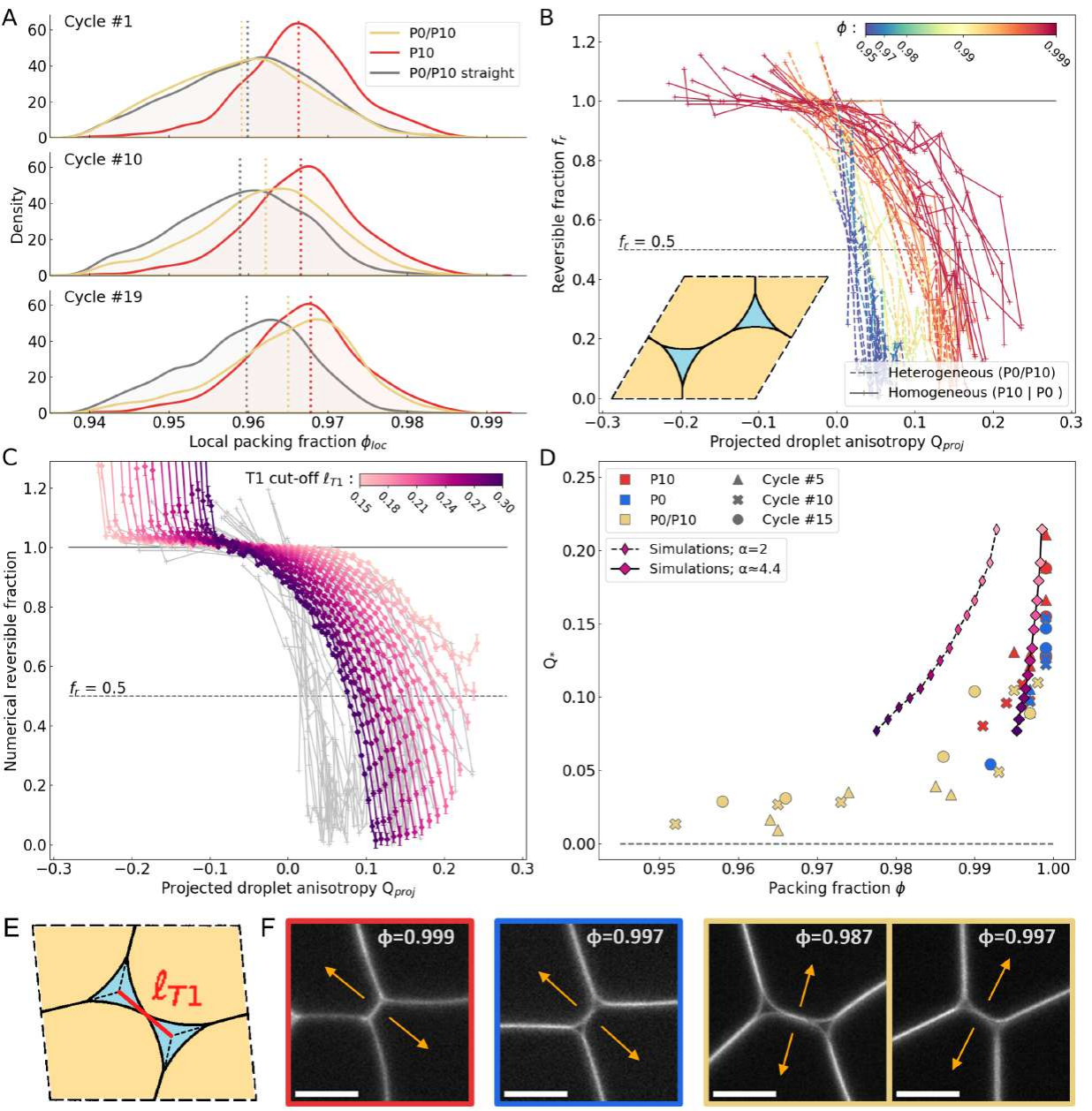}
\caption{
(A) Distribution of the packing fraction in static acquisitions for different experimental conditions and undulations in the channel. Heterogeneous emulsions (yellow, P0/P10) exhibit an overall shift towards higher values of packing fraction along the oscillatory channel (from top: undulation 1, to bottom: undulation 19), while heterogeneous emulsions in the straight channels (grey) start at the same value without exhibiting any evolution. The distributions of packing fraction for homogeneous emulsions in undulated channels (red) also remain constant across undulations. Vertical doted lines represent the mean value of each distribution. (B) Reversible fraction as a function of the projected droplet shape elongation $Q_\mathrm{proj}$, colored according to the packing fraction $\phi$ measured for each movie. Dashed lines: heterogeneous P0/P10 emulsions, solid lines: homogeneous P0 and P10 emulsions. Inset: sketch of two tricellular junctions from which the packing fraction is estimated in the movies (adapted from \cite{Princen1983}). (C) Reversible fraction computed from vertex model simulations at different T1 cut-off length $\ell_{\rm T1}$. In gray, we reproduce the experimental curves from panel B. (D) Evolution of $Q_\ast$ as a function of the packing fraction $\phi$ for experiments and numerical simulations. In the case of numerical simulations, the T1 cut-off lengths $\ell_{\rm T1}$ have been converted into packing fractions using Eq.~(\ref{eq:phi-lT1}), using either Princen's prefactor $\alpha_0=2$ or $\alpha\approx4.4$, obtained through squared error minimization (see SI). (E) Representation of the T1 cutoff length $\ell_{\rm T1}$, defined at the moment when two tricellular junctions meet and are about to merge, thus triggering the T1 event. $\ell_{\rm T1}$ is defined as the distance between the two tricellular junctions' centers in this geometry (adapted from \cite{Princen1983}). (F) Last snapshot of the droplet-droplet interface before a T1 event for homogeneous (P10, red border and P0, blue border) and heterogeneous emulsions (P0/P10, yellow borders), 50~ms before the contact has disappeared (i.e. before the next snapshot). Scale bar is $10\,\micro\meter$, arrows represent the direction of the T1 (i.e. of the separation between previously contacting interfaces).}
\label{fig:Fig4}
\end{figure*}

Note that we also observe values of $f_r>1$ for the smallest $Q_\mathrm{proj}<0$ (Fig.~\ref{fig:Fig3}D). By definition, a value of $f_r>1$ means that droplet shape elongation in these cases is even larger than the shear strain, which essentially implies that T1 transitions occur \emph{perpendicular} to the external shear direction.
We make very similar observations in simulations of our emulsions (see below -- Fig.~\ref{fig:Fig4}C). 
In experiments and simulations, these perpendicular T1 transitions occur for negative $Q_\mathrm{proj}$ right after the reversal of the shear direction. We believe they represent regions that were close to undergoing T1s before the shear reversal, but are only triggered right after the shear reversal. 
The existence of these T1 transitions indicates a refinement of the simple picture given above, which suggests that there should be no T1 transitions for $Q_\mathrm{proj}<0$.

So far, we have hypothesized that $f_r$ depends only on the local projected cell shape, $Q_\mathrm{proj}$. Yet, there are also other possibilities, e.g.\ $f_r$ could additionally depend on other aspects of the local emulsion packing structure. Moreover, $f_r$ could also be affected by non-local effects, e.g.\ strain steps elicited by T1 transition in other regions of the emulsion that are propagated through long-range elastic interactions \cite{Nicolas2018a}. 
If $f_r$ depended mostly only on the local $Q_\mathrm{proj}$, then $f_r(Q_\mathrm{proj})$ should play the role of a \emph{constitutive relation} and it should be possible to predict the measured $Q_{xx}(x)$ curves knowing only $f_r(Q_\mathrm{proj})$ and the applied shear protocol (see Methods and SI).
If this hypothesis was wrong, and other dependencies of $f_r$ also played an important role, the reconstruction of $Q_{xx}(x)$ based on $f_r(Q_\mathrm{proj})$ should fail.
In Fig.~\ref{fig:shape}C, we compare the $Q_{xx}(x)$ curves predicted based on $f_r(Q_\mathrm{proj})$ (dashed lines) to the measured ones (circles, diamonds and pentagons).
We find good agreements between reconstructed and measured $Q_{xx}(x)$ for most of the emulsions.
Thus, a single dependency of $f_r$ on $Q_\mathrm{proj}$ alone captures most of the observed $Q_{xx}(x)$ behavior. This includes specifically the shift in $\Delta Q_{xx}$ across shear cycles in the P0/P10 emulsions.
This strongly suggests that the observed shift in droplet asphericity in heterogeneous emulsions is due to a shift in the $f_r(Q_\mathrm{proj})$ curves towards higher $Q_\mathrm{proj}$ values.

\subsection*{Simulations exclude heterogeneous interface tensions or energy barriers to T1 transitions as reasons for the observed shift}
\label{section:vertex model framework}
To study possible reasons for the shift in the yielding behavior, we carried out simulations of 2D vertex models (see Methods) \cite{Alt2017}. Vertex models describe emulsions, foams, and biological tissues as polygonal tilings, where each polygon corresponds to one droplet.
We subject the vertex model to an oscillatory pure shear, where after each shear step, we quasi-statically minimize the system energy stemming from the droplet-droplet adhesion. Droplets are allowed to rearrange if a droplet-droplet contact shrinks below a length given by a T1 cutoff parameter $\ell_\mathrm{T1}$, defined in terms of the average droplet area.
Using a mixture of two kinds of droplets, we studied whether a shift in the yielding behavior could be created by adhesion heterogeneities, where we tested two different ways of representing adhesion in our simulations.

First, we tested whether adhesion heterogeneities could create substantially different droplet-droplet interface tensions, and whether this could create the observed shift.
To experimentally obtain the interface tension ratio in our P0/P10 emulsions, we measured the contact angles between interfaces at three-droplet junctions and found an interface tensions ratio of \emph{at most} $1:0.68$ (see SI \& Fig.~S11).
Using this maximally possible tension ratio, we ran simulations with heterogeneous interface tensions, and compared with the case of homogeneous interface tensions. 
Indeed, there was a shift across subsequent shear cycles in the yielding behavior of the heterogeneous in-silico emulsions (Fig.~S12D), like in the experiments. Yet, the amount of shift was small, and we observe the same amount of shift in the homogeneous in-silico emulsions (Fig.~S12C), in contrast to the experiments. 
This is due to a known effect, where under oscillatory shear the structure of the emulsion shows an initial transient evolution \cite{kabla2007quasistatic}.
Furthermore, comparing Figs.~\ref{fig:Fig3}D, S4 right, and S12A,B, we find that unmixing and yielding behavior of heterogeneous in-silico emulsions is inconsistent with our experimental data.
In short, in-silico emulsions unmix much faster and/or yield less easily than our experimental emulsions.
Taken together, the modulation of interface tensions by adhesion can not explain the difference in shift that we observe between homogeneous and heterogeneous emulsions.

Second, given that adhesion also sets a barrier for droplet-droplet detachment, we tested whether heterogeneities in T1 transition barriers could create the observed shift.
In the simulations, we describe the T1 transition barrier by the T1 cut-off length $\ell_{\rm T1}$.
The most adhesive emulsions would thus be associated to smaller values of $\ell_{\rm T1}$ as compared to less adhesive ones. 
In this scenario, the homogeneous, less adhesive in-silico emulsions yield more easily than the more adhesive ones (Fig.~S12F), which is in contradiction with our experimental observations (Fig.~\ref{fig:Fig3}D).
Moreover, the reversible fraction function of the heterogeneous in-silico emulsion lies in between the two homogeneous cases (Fig.~S12F), again in contrast to our experimental observations.
Taken together, homogeneous emulsions with a lower barrier for T1 transitions will yield more easily, suggesting that T1 barrier heterogeneity can not play a major role in our emulsions.

\subsection*{Progressive compaction explains the shift in yielding behavior}
To understand what else might cause the observed shift in yielding behavior, we noted that, in the simulations, the T1 cutoff $\ell_{\rm T1}$ strongly affected the yielding behavior (Fig.~S12).
While this parameter is not substantially tuned by adhesion (see previous paragraph), it is expected that the barrier towards T1 transitions depends on the packing fraction $\phi$, i.e.\ the area fraction of oil within the emulsion. Indeed, for a lower packing fraction, one would expect the droplets to rearrange more easily.
To relate the T1 cutoff and packing fraction, Princen considered the triangular regions of continuous phase where three droplets meet (blue regions in Fig.~\ref{fig:Fig4}B inset) \cite{Princen1983}.
He found:
\begin{equation}
	\ell_{\rm T1} = \frac{\alpha \sqrt{1-\phi}}{\sqrt{\rho}},\label{eq:phi-lT1}
\end{equation}
where $\rho$ is the average area number density of the droplets (in the simulations, $\rho=1$). Crucially, Princen assumed that droplets would rearrange exactly \emph{at the moment when two triangular regions meet} (Fig.~\ref{fig:Fig4}E). For ordered packings of non-adhesive, monodisperse, hexagonally packed droplets, this implies $\alpha=\alpha_0\equiv 2/\sqrt{3(2\sqrt{3}-\pi)}\approx 2.0$ \cite{Princen1983}.

We therefore measured the local packing fractions in our emulsions. Indeed, the heterogeneous P0/P10 emulsions display an increase in packing fraction across subsequent shear cycles (see yellow curves in Fig.~\ref{fig:Fig4}A), while the packing fraction distribution is maintained throughout the channel for homogeneous emulsions (red curves). Furthermore, heterogeneous P0/P10 emulsions that are not exposed to cyclic shear also show a roughly constant packing fraction distribution (grey curves). 
This suggests that the adhesion differential modifies the flow properties of our emulsion such that the water phase can be progressively expelled under the cyclic shear, thus increasing its packing fraction.
Yet, this progressive compaction only occurs when both an adhesion differential and a repetitive strain are acting together.

To study the effect of the packing fraction on the yielding behavior, we color the reversible fraction curves from Fig.~\ref{fig:Fig3}D as a function of the packing fraction (Fig.~\ref{fig:Fig4}B).
Indeed, we find a clear trend where, as the packing fraction increases, emulsions yield less easily, i.e.\ the droplets need to be deformed up to higher $Q_\mathrm{proj}$ until yielding. 
We compare these results to vertex model simulations, where we vary the T1 cutoff from $\ell_{\rm T1}=0.15$ to $0.3$ (Fig.~\ref{fig:Fig4}C). We note that the $f_r$ curves of our in-silico emulsions have a similar qualitative shape as our experimental curves. Furthermore, the in-silico emulsions yield less easily as the T1 cutoff is decreased.

To compare experiments and simulations more directly, we quantify the droplet shape $Q_\ast$ where the reversible fraction is $f_r(Q_\mathrm{proj}=Q_\ast)=0.5$ (see gray dashed lines in Fig.~\ref{fig:Fig4}B,C, SI). In Fig.~\ref{fig:Fig4}D, we plot $Q_\ast$ as a function of the packing fraction $\phi$. 
We first note that when plotting our experimental data against $\phi$, strikingly, our data almost collapse to a master curve, which means that there is no significant effect of adhesion hierarchy on yielding (blue vs.\ red vs.\ yellow data points).
This observation is a central result of our work.
We compare this experimental data to simulations (shades of magenta), where the packing fraction $\phi$ is computed from the T1 cutoff $\ell_{\rm T1}$ using Eq.~(\ref{eq:phi-lT1}) with different values of $\alpha$.
For the packing fraction range $\phi\gtrsim0.97\dots0.99$, covered by our experiments, the idealized value $\alpha_0\approx 2.0$ captures qualitatively the experimentally observed trend but does not match quantitatively. Our experimental data are actually best fitted with $\alpha\sim 4.4$ (see SI). 
This indicates that the simple criterion from Princen is insufficient to capture our experimental findings, and yielding can occur already before the two continuous-phase triangular regions meet (see Fig.~\ref{fig:Fig4}E). 
Note that the vertex model simulations do not capture the experimental trend at low packing fractions ($\phi \leq 0.98$), i.e.\ high T1 cutoff length, which is due to the fact that vertex models are inherently not designed to model wet foams (see Methods).

To independently assess whether T1 fusions may occur before triangular regions touch, we created snapshots of droplet-droplet interfaces \emph{right} before a T1 transition is about to happen (Fig.~\ref{fig:Fig4}F). In all cases, and for different kinds of emulsions, we find that droplet-droplet interfaces still have a length on the order of the size of the triangular regions right before the T1 transitions are triggered, quite different from the scenario assumed by Princen (Fig.~\ref{fig:Fig4}E). Because of the time resolution of our movies, these observations cannot be considered as a quantitative measurement of $\ell_{\rm T1}$. Nevertheless, they are in qualitative agreement with our findings in Fig.~\ref{fig:Fig4}D.
This indicates that there could be a -- so far unknown -- instability through which droplet-droplet interfaces may already undergo a fast collapse at a finite length.

\section{Discussion}
Here we studied the impact of adhesion architecture on the mechanical properties of biomimetic emulsions.
To do so, we compared emulsions composed of a single droplet type, i.e.\ displaying an homogeneous adhesion, with emulsions composed of two distinct droplet populations, each with their own adhesion strength. 
We applied cyclic shear to these emulsions and developed a geometric framework to characterize their elastoplastic properties.
We showed that the emulsions with an adhesion differential changed their yielding behavior across shear cycles.
Comparing this approach with vertex model simulations revealed that this shift in yielding behavior was due to a progressive compaction.

It has long been known that viscosity, stiffness, and yielding behavior of particulate matter are closely linked to the packing fraction \cite{Raufaste2007,Forterre2008a,Liu2010,Das2023}.
Yet, here we demonstrated that for emulsions this relation can be quantified using imaging data only.
We extended the geometric formalism of Ref.~\cite{Merkel2017} to the quasi-static limit by defining a reversible fraction function $f_r$, which corresponds to the fraction of shear created by droplet shape changes alone.
We showed that the reversible fraction $f_r$ depends mostly only on the local projected droplet shape, $Q_\mathrm{proj}$, by reconstructing the observed droplet shape changes from the function $f_r(Q_\mathrm{proj})$.
This means that $f_r(Q_\mathrm{proj})$ can be considered as a constitutive relation describing the yielding of the emulsion. 
While the stress tensor does not explicitly appear, it can be directly computed from the droplet shape tensor $\bm{Q}$ knowing the interface tensions \cite{Batchelor1970,Weaire1999,Kraynik2003,kabla2007quasistatic}.
Our approach is quite general and relies only on the assumption of quasi-staticity. It can thus be used to quantify yielding behavior of many foams or emulsions but also biological tissues, without the need of explicit force measurements. 
Our approach could therefore also prove powerful to study morphogenetic processes that rely on transitions in tissue mechanics.

Future work can further refine the reversible fraction function $f_r(Q_\mathrm{proj})$. 
One extension could account for the observed values of $f_r>1$ both in experimental and in-silico emulsions. These values likely stem from T1 transitions triggered right after the shear reversals.
Another extension could account for non-local interactions. Specifically, in some of the data, there were indications of T1 transitions that have been triggered by other T1 transitions through long-range elastic deformations \cite{Nicolas2018a}. 
We further considered the system essentially as one-dimensional. Yet, we observed small heterogeneities of the flows with respect to the channel height (see SI). In future work, it should be possible to predict the 2D flows using a single 2D constitutive relation akin to $f_r(Q_\mathrm{proj})$.
Finally, it will be interesting to link our work to related ideas, e.g.\ studying the density of weak spots in cellular materials including biological tissues \cite{Popovic2021}.

Based on the geometric analysis of the yielding behavior, vertex model simulations helped us test how important the effect of adhesion was on droplet-droplet interface tensions or energy barriers towards droplet rearrangements. We showed that any substantial effect of either would be inconsistent with our experimental observations.
While the vertex model actually describes completely dry emulsions, we were able to account for packing fractions lower than $100\%$ by varying the T1 cutoff length. Yet, this approach started to fail at packing fractions below $\sim 98\%$, corresponding to a T1 cutoff of $\sim 0.3$, essentially, both because of a high droplet area polydispersity, and because the geometry of the continuous-phase triangular regions is completely ignored in the simulations (see Methods for details).
In the future, smaller packing fractions will be accessed by explicitly accounting for the continuous phase in our simulations \cite{Kim2021a}.
Nevertheless, our approach allowed us to show that packing fraction dominated in determining the yielding behavior of our emulsions.

Our results also suggest a deviation from the long-standing assumption, initially by Princen \cite{Princen1983}, that in foams and emulsions, T1 transitions occur when two continuous-phase triangular regions meet. Specifically, we provide direct and indirect evidence that T1 transitions were already triggered much earlier in our emulsions.
We are not aware of any previous report of such a deviation from the Princen rule, and so far, it is unclear what mechanism creates it.
For instance, it could be due to either a static instability that leads to an earlier triggering of the T1. Another possibility could be that frictional forces play a role, as previously discussed in foams \cite{Durand2006, Biance2009}. While the large-scale dynamics in our emulsions was quasi-static, friction might still influence the behavior of individual T1 transitions.
Regardless of the precise mechanism, while today Princen-like rules are often used when studying cellular materials, e.g.\ as a criterion for T1 transitions in vertex models, our finding shows that such classical ideas may not be sufficient to describe emulsions, and possibly neither biological tissues.
It would thus be interesting to explore whether an early triggering of T1s can also be observed in other soft matter and biological systems.

Importantly, our work also provides new insight in the context of animal morphogenesis, where spatial gradients or temporal increases in compaction have been observed in several developing animals and correlated with corresponding tissue rigidification \cite{Turlier2015a,Petridou2021,Mongera2018,Michaut2025}.
Recent work on biological tissues suggests that compaction and rigidification can result from the presence of a higher adhesion overall \cite{Ray2025}, and that adhesion and compaction can be disentangled as control parameters, with adhesion being more important for rigidification \cite{Rustarazo-Calvo2025}.
In contrast, in our droplet emulsions, we observed \emph{progressive} compaction and rigidification, which only occurred for emulsions with an adhesion \emph{differential} and only under ongoing (cyclic shear) deformation. 
Using mass conservation, one can furthermore show that the flow speed of the continuous phase has to increase in magnitude across subsequent shear cycles (see SI).
In other words, the interplay of heterogeneous adhesion and cyclic shear may create an effective \emph{pumping} of the continuous phase with respect to the droplet phase, and preliminary data tracking the continuous phase suggests that, indeed, the continuous phase generally flows faster than the droplet phase.
While the precise origin of such a pumping mechanism remains to be elucidated, it might explain the emergence of compaction \cite{Petridou2021,Mongera2018,Michaut2025} or contribute to fluid flows \cite{Dumortier2019} in biological tissues. 
More generally, it highlights how cellular material rheology can be modulated through cellular packing, which is in turn driven by externally applied deformation.

\begin{acknowledgments}
\section{ACKNOWLEGMENTS}
We thank Clement Nizak and Rapha\"el Doineau for their help with the design of the microfluidic chip, as well as Jacques Fattaccioli and Heloise Uhl for the use of the membrane emulsifier. We thank Georges Debregeas, Benjamin Dollet and Alexandre Kabla for fruitful conversations regarding foam Physics. We also acknowledge funding from EMERGENCE(s) Ville de Paris. Finally, this work was granted access to the HPC resources of the SACADO MeSU platform at Sorbonne Universit\'e.

L.-L.P. and Q.G. designed experiments; Q.G. performed experiments; Q.G., M.B., M.M. analyzed data; M.M., M.B. and R.V. developed the model and performed simulations; Q.G., M.B., R.V., A.M.P., E.W., M.M. and L.-L.P. interpreted the results; M.M. and L.-L.P. wrote the manuscript.
\end{acknowledgments}

\appendix

\section*{APPENDIX}

\subsection{Emulsion Preparation}
\label{Emulsion Preparation}
All products are obtained from Sigma Aldrich, unless mentioned otherwise. 
We first prepare an oil-in-water emulsion using a pressure emulsifier (Internal Pressure Type, SPG Technology Co.). 
50~cSt silicone oil is emulsified through a SPG membrane (Shirasu Porous Glass, hyrdophilic, pore size$=10\,\micro\meter$ $\diameter$) in a 10~mM Sodium Dodecyl Sulfate solution (SDS). The obtained emulsion of oil droplets in water can be kept at room temperature for weeks.

The droplets are stabilized with phospholipids through the following procedure \cite{Zhang2017, Pinon2018}: 9~mg of egg L-$\alpha$-phosphatidylcholine (EPC) and 1~mg of DSPE-PEG(2000)-biotin are dried under Nitrogen and resuspended in 500~µl of Dimethyl Sulfoxide (DMSO). This solution is diluted with 4.5~ml of a 5~mM SDS aqueous buffer (5~mM SDS, 10~mM Tris, pH = 7.5) and sonicated for 30~min at room temperature. 2~ml of creamed emulsion is then added to this solution, incubated overnight at 4~°C and washed in the morning with 250~ml of the 5mM SDS buffer in a separating funnel. We repeat this procedure once in a 1~mM SDS buffer (1~mM SDS buffer, 10~mM Tris). The emulsion is finally washed with 250~mL of the 1~mM SDS buffer and stored at 4~°C for several weeks. Average diameter $\langle D \rangle = 29.6\,\micro\meter$, polydispersity index: 13.06\%.

\subsection{Emulsion Functionalization}
All DNA sequences share the same structure: a biotin molecule, followed by a 49~bp sequence that is common to every strand and that we call the \textit{backbone}, and finally a palindromic \textit{sticky end} sequence whose length ranges from 0~bp to 14~bp (P0, P6, P10, P14). All sequences are detailed in Supplementary Information.

The chosen binding energies are low enough to allow the droplets to detach during rearrangements without pulling the lipids from the oil/water interface, which is verified by the fact that droplets keep their color integrity throughout all experiments. 

Each DNA sequence is prepared separately. First, the backbone is hybridized with a 48~bp complementary sequence (CS) in order to obtain a stiff double stranded spacer between the sticky end and the biotin anchor. By construction the \textit{backbone} and the \textit{sticky end} are therefore separated by a 1~bp spacer. To do so,  192~pmol of the desired sequence and 192~pmol of CS are dissolved in 200~µl of filtered TS-buffer (1~mM SDS, 10~mM Tris pH= 7.5, 10~mM of NaCl) in a 0.5~ml DNA Lo-Bind tube (Eppendorf), and incubated at 35°C during 30~min.
We then add 5.7~µl of 1~mg/ml streptavidin, Alexa Fluor conjugate (ThermoFisher Scientific) to the DNA sequences and incubate the solution for 30~min in the dark at 35°C. 
An excess of DNA is used at this step so that all the streptavidins carry at least one DNA strand. This prevents the presence of free streptavidin in solution that could induce droplet adhesion through biotin-streptavidin-biotin bridges.
We use streptavidin, Alexa 594 for the palindrome with the longest sticky end (red droplets in all images) and streptavidin, Alexa 488 for the other sequence with a shorter sticky end (blue droplets). 
Finally, 100~µl of creamed emulsion, prepared as indicated in the previous section, are added to the DNA solution and incubated in the dark at 35°C for 1h. Gentle agitation is applied every 20~min in order to resuspend the emulsion. 

After this final incubation the droplets are rinsed 3 times with 200~µl of filtered TS-buffer and once with a filtered TS$_{gly}$-buffer (1~mM SDS, 10~mM Tris pH= 7.5, 10~mM of NaCl in a solution of 1:1 w:w glycerol:water). This buffer ensures a better match between the refractive indices of the oil and aqueous phases, which in turn facilitates imaging of the fluorescent droplet edges. Immediately before the experiment, 100~µl of each droplet population are mixed together in 1~ml of filtered TN30S-buffer (1~mM SDS, 10~mM Tris, 30~mM NaCl, 0.05~mg/ml $\beta$-casein from bovine milk, in a solution of 1:1 w:w glycerol:water, pH= 7.5).

\subsection{Experimental Set-Up}

The microfluidic channels are engineered as described in~\cite{Golovkova2020}. The whole channel is 30~µm high, which  confines the droplets in a 2D monolayer, and consists of three main parts. The first section of the channel is 315~µm wide and 1.65~mm long and is lined with ten 15~µm wide evacuation channels. These evacuation channels allow us to reach reproducible high packing fractions downstream. In the second area, the width of the channel presents 20 oscillations between 315~µm and 185~µm, imposing 20 shear cycles, with a periodicity of 420~µm. The last area contains a constriction that decreases over a length of 385~µm from a width of 315~µm to 25~µm. Afterwards the channel keeps its width of 25~µm for 583~µm before the final outlet.

Once mounted, the channel is passivated by flowing a solution of casein at $0.25~mg/ml$ for an hour ($\beta$-casein from bovine milk), before injecting the emulsion in the channel using a pressure pump (MFCS-8C, Fluigent). The microfluidic device is maintained at a temperature ranging between 18 and 20°C thanks to a custom microscope stage made of PMMA (see Supplementary Information), in which we circulate cooled water from a thermo-regulated bath (Cooling bath thermostat, CC-K6 Huber).

For static acquisition, once the emulsion is packed inside the channel, we repeat the following protocols for several set of acquisitions: the emulsion is let to flow at low speed for ten minutes, the flow is then stopped by progressive decrease of the applied pressure. We image the droplets at all undulations through spinning disk confocal microscopy using a 20x objective (Spinning Disk Xlight V2, Gataca systems) and finally resume the flow to evacuate the previous emulsion.

For dynamic acquisitions, we acquire movies of 1000 images at $20\,\hertz$ in a single undulation of the channel. We image three undulations (\#5, \#10 and \#15) and vary the applied pressure, i.e.\ the flow velocity. The average flow velocities range from $\sim 8$ to $\sim50\,\micro\meter/\second$. It thus takes the droplets between $\sim$8.4 to $\sim$52.5~s to travel across one undulation, which is sufficient to initiate new adhesion between droplets~\cite{Bourouina2011a}. Note that the lower bound of flow velocity roughly corresponds to the one used to flow the emulsion in between two static acquisitions.

\subsection{Image analysis}

We analyze separately the channels corresponding to each droplet population. First, images are segmented using Ilastik~\cite{Berg2019a} in order to separate the fluorescent contour of the droplets from the background. Using a homemade Fiji routine and the segmented images, we create a mask for both channels in order to identify the droplets and compute a surface Voronoi tessellation of the whole packing (in such tesselation, the whole contour of the droplet is considered to be the seed of the Voronoi cell, which automatically takes in account the polydispersity of the system). We exclude Voronoi cells on the edge of the image, i.e.\ we exclude droplets on the edge of the channel and partial droplet images. These steps are exemplified in the attached SI. 
The rest of the analysis is performed using the Sci-kit image Python module. For each image, we start by labeling the Voronoi tessellation and the combination of red/green masks of the droplets. Each droplet is associated to its Voronoi cell and its color. We compute the local packing fraction $\phi_{loc}$ as the ratio between the droplet area and its Voronoi cell area, note that to measure the packing fraction in the dynamic acquisition, we evaluated $\phi$ from the area of the triangles at tri-cellular junctions which corresponds to a similar measurement (Fig.~S9B). The positions of the droplets are extracted from the centroid coordinates of a fitted ellipse. In addition, we extract the perimeter $p$ and surface $a$ of each droplet to calculate its corresponding shape factor $\mathcal{A}~=~p^{2}/4\pi a$. 

From the segmented images, we identify inner vertices (i.e.\ vertices involving at least 3 droplets, thus excluding droplets along the channel border) in order to triangulate the network as in \cite{Merkel2017} (details in the SI). Triangles elongation  is characterized by the tensor $\bm{Q}$. The triangles and their connectivity are analyzed over time, allowing one to measure the shear contributions by droplet deformation and T1 transitions in order to compute the reversible fraction $f_r$.

\subsection{Prediction of droplet shape}
In Fig.~\ref{fig:shape}C, we use quantified reversible fraction curves, $f_r(Q_\mathrm{proj})$, to predict the droplet shapes $Q_{xx}(x)$ as the emulsion is pushed through the undulated channel.
To this end, we start in our 1D picture from the definition of the reversible fraction, which implies:
\begin{equation}
    \frac{\d Q_{xx}}{\d t} = f_r(Q_\mathrm{proj})\tilde{V}_{xx},
\end{equation}
where $\d Q_{xx}/\d t=\partial Q_{xx}/\partial t + \mathrm{v}_x \partial_x Q_{xx} $ with $\mathrm{v}_x$ being the $x$ component of the local velocity, and $Q_\mathrm{proj}=\mathrm{sgn}(\tilde{V}_{xx})\, Q_{xx}$.
Using stationarity, $\partial Q_{xx}/\partial t=0$, we thus obtain:
\begin{equation}
    \partial_x Q_{xx} = f_r\Big(\mathrm{sgn}(\tilde{V}_{xx})\, Q_{xx}\Big) \frac{\tilde{V}_{xx}}{\mathrm{v}_x}.
\end{equation}
With incompressibility, we have $\tilde{V}_{xx}=\partial_x \mathrm{v}_x$. Because the $\log$ function is monotonously increasing, $\mathrm{sgn}(\tilde{V}_{xx})=\mathrm{sgn}(\partial_x\log{\mathrm{v}_x})$, implying:
\begin{equation}
    \partial_x Q_{xx} = f_r\Big(\mathrm{sgn}\big[\partial_x\log{\mathrm{v}_x}\big]\, Q_{xx}\Big) \partial_x\log{\mathrm{v}_x}.
    \label{eq:reconstruction}
\end{equation}
If the velocity $\mathrm{v}_x$ was perfectly homogeneous along the width $h$ of the channel, we would have $h\mathrm{v}_x=\mathrm{const.}$ and thus $\partial_x\log{\mathrm{v}_x}=-\partial_x\log{h}$.
Yet, we observe that there are some variations of $\mathrm{v}_x$ with channel width and so we use a phenomenological fit to the measured $\mathrm{v}_x(x)$, which we then insert into Eq.~(\ref{eq:reconstruction}) (see SI for details).

To solve Eq.~(\ref{eq:reconstruction}), we need to be able to compute $f_r$ for any value of $Q_\mathrm{proj}=\mathrm{sgn}\big[\partial_x\log{\mathrm{v}_x}\big]\, Q_{xx}$.
Yet, for the experimentally determined $f_r$ curves, we only have discrete $(f_r, Q_\mathrm{proj})$ data points.
We thus fit these data points to the phenomenological function $f_r=1-0.5\,\exp{([Q_\mathrm{proj}-Q_\ast]/\lambda)}$ with fit parameters $Q_\ast$ and $\lambda$. This fit function is then inserted into Eq.~(\ref{eq:reconstruction}).

Finally, we integrate Eq.~(\ref{eq:reconstruction}) using an explicit Euler method with a spatial stepping of $\Delta x=1\,\micro\meter$. We ran the integration for a number of cycles corresponding to the cycle of the experimental observation (i.e.\ 5 or 15) and compare in Fig.~\ref{fig:shape}C only the last cycle to the experimentally determined $Q_{xx}(x)$ data.

\subsection{Vertex model simulations}
Our vertex model describes a 100\% dense packing of $N=400$ polygonal droplets, where the degrees of freedom are the positions of the polygon corners, called vertices. The mechanics of our model is defined by the following energy functional:
\begin{equation}
    E = \frac{1}{2}\sum_{i=1}^N{k_A(A_i-A_{0i})^2} + \sum_{\langle i,j\rangle}{\lambda_{ij}\ell_{ij}}.
    \label{eq:vm energy}
\end{equation}
Here, the first sum is over all droplets $i$, where $k_A$ is parameter denoting an area elastic modulus, $A_i$ denotes the actual droplet area, and $A_{0i}$ is a droplet-dependent target area. 
The second sum is over all pairs $\langle i,j\rangle$ of neighboring droplets. These are unordered pairs, i.e.\ each pair $\langle i,j\rangle\equiv\langle j,i\rangle$ appears only once in the sum.
In this sum, $\lambda_{ij}$ denotes an effective interface tension and $\ell_{ij}$ is the interface length. 

We use periodic boundary conditions, where we apply varying dimensions $L_x\times L_y$, but with a constant total area $L_xL_y=N$, i.e.\ each cell has on average an area of 1 available to it. 
We initialize our in-silico emulsion as the Voronoi tessellation of a random point pattern.
Afterwards, we apply cyclic pure shear by setting $L_x=\sqrt{N}\,e^\gamma$, where the shear strain $\gamma$ varies between $\gamma=-\hat\gamma$ and $\gamma=+\hat\gamma$ with $\hat\gamma=0.27$ and constant shear step $\Delta\gamma=\hat\gamma/100=0.0027$.
The strain amplitude is taken from the experiments, computed as: $\hat\gamma= \log{ ( w/[w-2\Delta])}/2 \approx 0.266$, where $w=315\,\micro\meter$ and $\Delta=65\,\micro\meter$ (compare Fig.~\ref{figure1}C).
The value of $\hat\gamma=0.27$ was used in the all simulations (Fig.~\ref{fig:Fig4}C and Figure~S12). 
All results (Fig.~\ref{fig:Fig4}C and S12) were averaged over 50 to 150 simulation runs with different random realizations of the initial conditions.

After each shear step, we quasi-statically minimize the energy in Eq.~(\ref{eq:vm energy}) using a custom Conjugate-Gradient algorithm.
During the energy minimization, we test whether the interface length between any two droplets $i$ and $j$ is below the T1 cutoff $\ell_{\mathrm{T1};ij}$. If this is the case, we fuse the two vertices into a many-fold vertex (i.e.\ a vertex where more than four droplets meet).
Moreover, we also test if forces applied to a many-fold vertex allow for the stable formation of a new interface, in which case we split it into a new interface with length $1.5\ell_{\mathrm{T1};ij}$.

We use the following parameter values.
We set $k_A=100$ to enforce an effective incompressibility of the droplets. To match the experimental polydispersity, we draw for each droplet $i$ the target area $A_{0i}$ from a Gaussian with a standard deviation of $0.4$ and average of $1$, where we impose a lower cutoff of $0.3$ to prevent droplets from disappearing.
We impose a number ratio of $1:1$ between P0 and P10 droplets, corresponding to the approximate experimental value (Fig.~S4 left).
For simulations with a homogeneous interface tension, we always use $\lambda_{ij}=1$ (Fig.~\ref{fig:Fig4}C and Fig.~S12C,E,F).
For simulations with heterogeneous interface tensions (Fig.~S12A,B,D), we use:
\begin{equation}
    \lambda_{ij} = \begin{cases}
        0.68 & \text{if $i$ and $j$ are both P10}\\
        1 & \text{if $i$ or $j$ is P0.}
    \end{cases}
\end{equation}
Similarly, for the simulations with heterogeneous T1 cutoffs (Fig.~S12E,F), we set $\ell_{\mathrm{T1};ij}$ to $0.1$ if both $i$ and $j$ are P10, and to $0.25$ if at least one of both is P0.

We realized that in our vertex model simulations, for large $\ell_{\rm T1}$ the energy minimization became problematic in the sense that (1) many-fold vertices tended to form (i.e.\ vertices that abut \emph{more than} three cells), and (2) there were always new T1 events occurring during the minimization. To accommodate for problem (1), we ran simulations where many-fold vertices were allowed. We tried to also handle (2) by allowing only a given maximal number of 3 T1 transitions on any given edge or vertex. Yet, this resulted in pathological configurations (e.g.\ many triangular cells, many concave cells), which is likely due to T1s also getting ``stuck'' this way, preventing a proper minimization.
That there will be problems at large $\ell_{\rm T1}$ becomes clear already from the fact that we have a substantial area polydispersity with the minimal area cutoff at 0.3. Indeed, e.g.\ the side length of a hexagon with an area of $0.3$ is $\sqrt{2\times0.3/3\sqrt{3}}\approx 0.34$. Thus, already from this very optimistic estimate, one would expect problems related to ``never-ending'' T1 transitions at least for $\ell_{\rm T1}\gtrsim0.34$.

\bibliography{Differential}

\end{document}